# Does religiosity influence corporate greenwashing behavior?

Mathieu Gomes[a,*]

Sylvain Marsat[a]

Jonathan Peillex[b]

Guillaume Pijourlet[a]

[a]Université Clermont Auvergne, CleRMa, 11 bd Charles de Gaulle, 63000 Clermont-Ferrand, France.

[b]ICD International Business School, 12 rue Alexandre Parodi, 75010 Paris, France.

## Abstract

We analyze the influence of religious social norms on corporate greenwashing behavior. Specifically, we focus on a specific form of greenwashing: selective disclosure. Using a large sample of US firms between 2005 and 2019, we show that firms located in counties where religious adherence is high are less likely to engage in greenwashing. We also find that a stronger religious adherence within the county in which a company is located reduces the magnitude of greenwashing, when observed. We further analyze the mechanism underlying this relationship and show that religious adherence impacts greenwashing behaviors through the channel of risk aversion. A comprehensive set of robustness tests aimed at addressing potential endogeneity concerns confirms that religion is a relevant driver of corporate greenwashing behavior.

# 1. Introduction

The impact of religion on various economic outcomes has been the subject of many academic studies (Barro and McCleary, 2003; Guiso et al., 2003, Iannaccone, 1998; Lehrer, 2004; among others). Interestingly, researchers have only recently started to look at the influence of religious adherence on corporate decision making and especially on unethical manipulative corporate behavior (Hilary and Hui, 2009; Dyreng et al., 2012; McGuire et al., 2012; Kanagaretnam et al., 2015; Kirchmaier et al., 2018; Leventis et al., 2018; Cai et al., 2019; Harjoto et al., 2019; Chantziaras et al., 2020; Abdelsalam et al., 2021; Terzani et al., 2021).

Despite this increased interest, no research has been conducted thus far to study the relationship between religious adherence and corporate greenwashing behaviors. Nonetheless, greenwashing practices are typically an example of manipulative behavior (Delmas and Burbano, 2011; Boncinelli et al., 2023), consisting in giving a distorted image of the environmental efforts made by the company in order to influence consumer behavior (Meisinger, 2022, Santa and Drews, 2023). It has however been shown that religiosity has a negative influence on corporate unethical behavior such as earnings management (Abdelsalam et al., 2021; Cai et al., 2019; Kanagaretnam et al., 2015; McGuire et al., 2012; Dyreng et al., 2012).

In this paper, we focus on one type of organizational greenwashing—the selective environmental disclosure—which consists of disclosing "positive environmental actions while concealing negative ones to create a misleadingly positive impression of overall environmental performance" (Marquis et al., 2016). Selective disclosure can be seen as a form of greenwashing insofar as some information is voluntarily not disclosed to create a positive impression (Marquis et al., 2016). Because greenwashing may produce negative effects on consumers and investors' confidence (Arouri et al., 2021; Zhang et al., 2018) and prevent the development of sustainable markets (Boncinelli et al., 2023), it is crucial to understand the drivers of such behavior.

According to social norm theory, individuals adopt the behavior and values of the social group with which they associate, and conform to the dominant social norms of the group (Dyreng et al., 2012; Kohlberg, 1984; Sunstein, 1996; McGuire et al., 2012). Neo-institutionalist theory also suggests that organizations adopt these social norms to maintain their legitimacy in society, and thus ensure their survival (Meyer & Rowan 1977; DiMaggio & Powell 1983). Two major social norms promoted by religions may influence corporate decisions and lead us to hypothesize a negative impact of religious adherence on greenwashing. First, religions are unanimously opposed to the principle of lying and manipulation of others (Callen and Fang 2015; Kanagaretnam et al., 2015). This would be in line with promoting honesty, avoiding manipulative corporate behaviors and condemning actions aimed at hiding bad news from investors (Callen and Fang, 2015). Second, the literature clearly shows that religious people are more risk averse than others (Miller and Hoffmann, 1995; Dohmen et al., 2011; Noussair et al., 2013; Diaz, 2000; Iannaccone, 1998) and that religious adherence is associated with risk-averse behaviors while irreligiosity tends to be associated with risk-taking characteristics (Miller and Hoffmann, 1995). To the extent that greenwashing can pose a serious reputational risk if uncovered, even more so if it contradicts the prevailing behaviors of the social group with which firm managers interact (Chircop et al., 2017), we thus expect that firms headquartered in counties with high religious adherence will be less likely to greenwash.

In our study, we investigate the relationship between the religious adherence of the population in the county where the firm is headquartered and corporate greenwashing behavior. Specifically, we study a sample of 601 US firms over the 2005-2019 period, yielding 6,650 firm-year observations. We show that a stronger religious adherence within the county in which a company is located both reduces the likelihood to engage in and the magnitude of greenwashing. We also reveal that religious adherence impacts greenwashing behaviors through the channel of risk aversion.

Our paper's contribution is threefold. First, we contribute to the literature linking religious adherence to corporate decision making (Abdelsalam et al., 2021; Chantziaras et al., 2020; Dyreng et al., 2012; Hilary and Hui, 2009; Leventis et al., 2018; Li et al., 2023) by being the first to study the impact of religious adherence on corporate greenwashing behaviors. Second, we contribute to the greenwashing literature by investigating the influence of religious social norms on selective disclosure practices. Delmas and Burbano (2011) mention different types of greenwashing drivers: market external, organizational and individual psychological drivers. Thus, the existing literature has already identified several important market external drivers of greenwashing strategies. For instance, the influence of stakeholder pressure (Marquis et al., 2016; Testa et al. 2018), competitive pressure (Arouri et al. 2021), or public attention (Pope et al., 2023) have already been documented in the literature. However, to our knowledge, little is known about the effect of social norms on greenwashing, even though some authors cite them as possible determinants of greenwashing strategies (Lyon and Montgomery, 2015). Our results suggest that the normative social environment in which managers operate and especially religious social norms are among the drivers of greenwashing behaviors. Third, we analyze the mechanisms through which the influence of religiosity on greenwashing materializes and evidence that the risk-aversion channel explains the relationship. This result allows us to gain a better understanding of the determinants of selective disclosure, and to empirically demonstrate that corporate managers identify such behavior as a source of risk for firms. Our results have important implications at a time where firms around the world seem to deploy great efforts to "green" themselves. Indeed, if religious social norms affect firms' greenwashing behavior, then, understanding how these norms affect managers may help regulators, contract designers, and financial market participants to shape appropriate standards (Sunder, 2005).

The rest of the paper is organized as follows: Section 2 presents the literature and theoretical development. The data and methodology are presented in Section 3. Section 4 features the empirical analyses and discusses results. Section 5 concludes.

## 2. Literature review and theoretical development

2.1. Religious adherence and firms' decisions

Religion has a significant impact on decision-making (e.g., Smith, 1790; Weber, 1905). For instance, researchers have shown that religious beliefs affect individuals' personalities which in turn influence economic behaviors and ultimately economic growth (Barro and McCleary, 2003; Guiso et al., 2003). At the individual level, numerous studies have shown that religiosity, defined as "the degree to which an individual adheres to the values, beliefs, and practices promulgated by religion" (Leventis et al., 2018), influences how individuals evaluate the costs and benefits of certain actions such as marital status, education, or health-related behaviors, among others (Iannaccone, 1998; Lehrer, 2004). In a business context, religiosity appears to be linked with less acceptance of unethical decisions (Longenecker et al. 2004) and higher ethical judgments (Vitell, 2009; Walker et al., 2012).

This link between religiosity and economic decision-making at an individual level has led researchers to assess the potential impact of religious practice on decisions made at the firm-level. Indeed, the fact that managers' characteristics largely determine corporate decisions and performance has been largely documented (Hambrick and Mason, 1984). For instance, political preferences (Bhandari and Golden, 2021; Hutton et al., 2014; Bhandari et al., 2020; Francis et al., 2016; Elnahas and Kim, 2017), psychological attitudes such as overconfidence (Hirshleifer et al., 2012; Malmendier et al., 2011) or optimism (Graham et al., 2013), retirement preferences (Jenter and Lewellen, 2015) or personal leverage choices (Cronqvist et al., 2012) seem to influence firms' decisions.

Religion is a strong mechanism promoting social norms which are likely to influence managers' behaviors (Dyreng et al., 2012; Abdelsalam et al., 2021). Indeed, social norm theory predicts that individuals tend to adopt the preferences of the social group to which they belong so as to avoid the sanctions associated with deviation from values and behaviors considered as socially acceptable (Dyreng et al., 2012; Kohlberg, 1984; Sunstein, 1996; McGuire et al., 2012). When a company is located in a geographical area where citizens have a strong religious practice, managers are more likely to interact with religious individuals. These interactions will confront managers with social norms and therefore values and moral standards promoted by religions and that must be respected (Dyreng et al., 2012). Social norms will emerge from these interactions between managers and religious people, as well as a system of sanctions in the case of deviations from these norms (Cialdini and Trost, 1998; Dyreng et al., 2012). Social norm theory teaches us that prevailing religious norms might therefore impact managers' decisions, whether or not they adhere to a religion. Companies may thus seek to align their behavior with the social norms of the social groups with which they are in contact, notably due to the need for the company to maintain its legitimacy (Chircop et al., 2017). Behaviors that deviate from these social norms can indeed be penalizing for the company, resulting in the loss of support from the relevant social groups. According to this view, religious adherence has been shown to influence on corporate risk-taking (Adhikari and Agrawal, 2016; Shu et al., 2012), corporate investments (Hilary and Hui, 2009; Khedmati et al., 2021), financing policies (Cai and Shi 2019), venture capital decisions (Chircop et al., 2020) or audit pricing (Leventis et al., 2018).

The role of religious values as a promoter of greater CSR (Harjoto & Rossi 2019; Chantziaras *et al.* 2020) and business ethics practices (Weaver and Agle 2002; McGuire et al. 2012) has also been established by previous works. Regular attendance at religious services may constitute an ethical teaching (McGuire et al. 2012) that can influence decision-making practices, particularly in a business context (Kennedy & Lawton 1998; Weaver & Agle 2002; Kanagaretnam *et al.* 2015; Leventis *et al.* 2018). For instance, religions seem to mitigate opportunistic managerial behaviors (Callen & Fang 2015) and lead to more ethical judgments (Walker *et al.* 2012). Thus, religious social norms seem to have a significant negative impact on unethical practices such as earnings management (Abdelsalam et al., 2021; Cai et al., 2019; Kanagaretnam et al., 2015; McGuire et al., 2012; Dyreng et al., 2012), and a positive effect on ethical behaviors like CSR reporting (Chantziaras et al., 2020; Terzani and Turzo, 2021) and CSR activities (Harjoto and Rossi, 2019).

2.2. Religious adherence and greenwashing

In this article, our aim is to examine whether religious adherence has an influence on environmental disclosure choices. In particular, our analysis focuses on determining whether social norms promoted by religions limit firms' propensity to engage in a particular type of greenwashing, namely selective environmental disclosure. In this section, we aim to set out the theoretical mechanisms by which social norms such as those induced by religiosity influence corporate greenwashing behavior.

Social norm theory suggests that managers adopt behaviors consistent with the social norms of the social groups with which they interact (Chircop et al., 2017). Neo-institutionalist theory also helps us to understand how contextual characteristics can determine corporate policies (Chircop et al. 2020) and especially environmental disclosure practices (Lyon and Montgomery 2015; Siano et al. 2017). This theory emphasizes the role of institutions in the emergence of conditions leading to the development of common organizational practices (DiMaggio and Powell 1983; DiMaggio and Powell 1991). This theory states that organizations tend to conform to social norms and expectations in order to maintain their legitimacy and thus ensure their survival (Meyer and Rowan 1977; DiMaggio and Powell 1983).

The literature has highlighted two social norms promoted by religions that can influence corporate decisions and lead us to hypothesize a negative impact of religious adherence on greenwashing. First, it has been established that religions promote an "anti-manipulative ethos" (Callen and Fang, 2015; Kirchmaier et al., 2018; Abdelsalam et al., 2021). Indeed, religions are unanimously opposed to the principle of lying and manipulation of others (Callen and Fang 2015; Kanagaretnam et al., 2015). This social norm is thus opposed to behaviors aimed at hiding bad news from investors (Callen and Fang, 2015). In line with this argument, the existing literature seems to point out that religiosity has a significant negative impact on corporate behaviors which are considered manipulative. For example, high religious adherence limits earnings management (Kanagaretnam et al., 2015; Cai and Shi, 2019; Dyreng et al., 2012; McGuire et al., 2012). Looking specifically at greenwashing strategies, we can also postulate that a high level of religious adherence reduces the propensity of managers to engage in misleading behavior such as selective disclosure. Selective disclosure is indeed a type of greenwashing that consists in diverting the attention of stakeholders (Siano et al. 2017) by disclosing unimportant environmental information to mislead them (Lyon and Montgomery, 2015; Marquis et al. 2016). The social norm of honesty promoted by religions may reduce the propensity of managers to minimize firms' environmental impact by unethically disclosing flattering indicators instead of indicators that more accurately describe reality.

Second, it has been shown that religious people are more risk averse than others (Miller and Hoffmann, 1995; Dohmen et al., 2011; Noussair et al., 2013; Diaz, 2000; Iannaccone, 1998). Religious adherence is associated with risk-averse behaviors while irreligiosity tends to be linked to risk-taking traits (Miller and Hoffmann, 1995). This is consistent with Pascal's wager: it can be considered as rational to believe in God, independently of his existence. In the event that God exists, a non-believer loses a lot and a believer gains a lot, whereas both the believer and the non-believer have nothing to lose in the event that God does not exist (Pascal, 1852). Miller and Hoffmann (1995) underline that this type of argument is in line with more recent works on risk management strategies. Accordingly, Diaz (2000) shows that religiosity influences both the frequency of gambling in Las Vegas and the amount played. Religious individuals are also less likely to accept risky lotteries (Hilary and Hui, 2009). At the firm-level, it is expected that companies headquartered in counties with high religiosity --and thus presumably feature a higher degree of risk aversion-- may be less likely to adopt risk-taking behaviors in order to conform to the prevailing behaviors of the social group with which managers interact (Chircop et al., 2017). It has been shown that companies located in counties with a high level of religious adherence tend to take less risk than others (Adhikari and Agrawal, 2016; Hilary and Hui, 2009; Chircop et al., 2017), and invest less (Hilary and Hui 2009).

In the context of studying greenwashing behaviors, one might expect managers of companies located in highly religious counties to adopt environmental disclosure choices that limit reputational or litigation risks. It has been shown that greenwashing can be detrimental to a firm's environmental legitimacy and may hurt investors' confidence (Arouri et al., 2021; Berrone et al., 2017). Also, the exposure to greenwashing seems to lead to a negative markets' reaction, generating negative abnormal returns (Du, 2015). All the arguments put forward in this section lead us to formulate the following hypothesis:

H$_1$: Firms located in counties with high religious adherence are, all else equal, less likely to greenwash.

## 3. Data and methodology

3.1. Sample selection

Following previous literature (Hilary and Hui, 2009; Chantziaras et al., 2020; Chircop et al., 2017; Chircop et al., 2020; Cai and Shi, 2019; Cai et al., 2019; Callen and Fang, 2015; Jiang et al., 2018; Dyreng et al., 2012; McGuire et al., 2012; Leventis et al., 2018) we restrain our analysis to US firms. Focusing our analysis on a single country allows us to control for heterogeneity in terms of institutional and legal characteristics, whose effect can be confounded with that of religiosity (Adhikari and Agrawal, 2016; Chantziaras et al., 2020). The US context is particularly interesting, given the importance that religion plays in the American society (Iannaccone, 1998; Leventis et al., 2018). The United States is also characterized by a high religious diversity, which makes it possible to avoid potential biases due to single religious denominations (Leventis et al., 2018).

Starting with the sample of US companies for which we have Trucost data regarding their environmental disclosures for the period 2005-2019, we match this information with religious adherence, financial and demographics data. We use religion adherence data from the American Religion Data Archive (ARDA), and retrieve financial data from the Refinitiv Datastream database. We also use data from the US Census bureau to construct our demographic variables. Firms with negative total assets, net sales, free cash flows or common equity were excluded from the sample. We also set aside firms for which we do not have at least five years of financial data. Our final sample contains 601 firms for the 2005-2019 period, resulting in 6,650 firm-year observations. Firms are located in 171 counties distributed within 40 states. Table 1 presents a detailed distribution of our sample observations across states, industries, and years. We observe that observations are pretty well balanced across years. The most represented industries are Industrials and Consumer discretionary, with 1,409 and 1,344 observations, respectively. The number of observations by state ranges from 8 for Maine to 881 for California.

[Insert Table 1 here]

3.2. Greenwashing measure

We derive our measure of greenwashing from the notion of selective disclosure, defined by Marquis et al. (2016) as "a symbolic strategy whereby firms reveal a subset of private information to create a misleadingly positive public impression." Selective disclosure is the form of greenwashing that has been the subject of the largest number of studies (Cho and Patten, 2007; Clarkson et al. 2008; Kim and Lyon 2011; Lyon and Montgomery, 2015; Philippe and Durand, 2011). We compute selective disclosure magnitude as the difference between two ratios provided by Trucost: the absolute disclosure ratio and the weighted disclosure ratio (Marquis et al., 2016; Arouri et al., 2021; Pope et al., 2023). These two ratios respectively aim to evaluate firms' environmental symbolic and substantive transparency. The difference between the absolute and weighted disclosure ratios allows us to capture the tendency of a company to publish its less harmful indicators (Marquis et al., 2016)[1].

The absolute disclosure ratio is calculated as the number of relevant environmental indicators disclosed by a company divided by the total number of relevant environmental indicators related to the company's activities. Trucost considers more than 464 industries, and considers for each of them the environmental indicators deemed relevant to the sectoral issues, based on "the US Toxic Release Inventory, the Federal Statistics Office of Germany (Destatis), the UK Environmental Accounts, Japan's Pollutant Release and Transfer Register, Australia's National Pollution Inventory, and Canada's National

---

[1] The very detailed and comprehensive online appendix built by Marquis et al. (2016) summarizes the procedure very well through examples. Interested readers may refer to it at:
https://dash.harvard.edu/bitstream/handle/1/27419737/marquis%2ctoffel%2czhou_greenwash.pdf?sequence=1&isAllowed=y

Pollutant Release Inventory" (Marquis et al., 2016). In short, the absolute disclosure ratio assesses the share of relevant environmental figures disclosed by a company.

The calculation of the weighted disclosure ratio considers the relative importance of the different environmental indicators in terms of environmental damage costs. The denominator of this ratio is the firm's total environmental damage costs estimated in dollars. This figure is calculated by first calculating the products between a "physical-factor-per-unit-revenue" (e.g., x tons of $CO_2$ emitted, y liter of water, etc. per dollar of activity) for a specific sector and the firm's sales in this sector for all sectors in which the company operates. Then, these estimates of emissions released and natural resources consumed by a firm during a year are summed and weighted by environmental damage cost factors estimated in dollar value (Marquis et al. 2016). The numerator of the weighted disclosure ratio is simply the sum of emissions released and natural resources consumed for which an information is disclosed, weighted by environmental damage cost factors. Hence, the weighted disclosure ratio allows to evaluate the share of a firm's environmental damage costs for which an information is disclosed (Marquis et al., 2016).

We assume that firms greenwash when their absolute disclosure ratio is superior to its weighted disclosure ratio, or in others words, when its symbolic transparency is higher than its substantive transparency (Arouri et al., 2021; Marquis et al., 2016). Indeed, in this case, the firm publishes relatively more indicators related to a small part of its total environmental impact than others. Based on both absolute and weighted disclosure ratios, we build two variables as proxies for greenwashing. Our first variable is *GREENWASHING_BIN*, which is a dummy variable taking the value one if the absolute disclosure ratio is above the weighted disclosure ratio, and zero otherwise. This variable evaluates whether a firm is engaged in greenwashing. Our second variable is *GREENWASHING_MAG*. This variable equals zero if the firm does not greenwash, i.e., if its absolute disclosure ratio is lower than its weighted disclosure ratio. If the absolute disclosure ratio is strictly above the weighted disclosure ratio, *GREENWASHING_MAG* equals the difference between the absolute disclosure ratio and the weighted disclosure ratio. This variable allows us to assess the magnitude of greenwashing, if any.

3.3. Religiosity measure

In line with the extant literature, we use data from the American Religion Data Archive (ARDA) to calculate our religiosity measure (Cai and Shi, 2019; Callen and Fang, 2015; Chantziaras et al., 2020; Dyreng et al., 2012; Hilary and Hui, 2009; Leventis et al., 2018). Specifically, the ARDA conducts surveys approximatively every 10 years and provides the number of religious adherents for each US county (Cai and Shi 2019; Berry-Stölzle and Irlbeck 2021). More precisely, we use survey data from the Religious Congregations Membership Study (RCMS), which contains the total number of congregations and adherents of more than 200 religious groups for each US county. In 2010, the 236 religious bodies covered by the survey included more than 150,000 members, or approximately 48.8% of the US population. Our proxy for religiosity (*RELIGIOSITY*) is calculated as the number of religious adherents for all denominations as reported by religious groups surveyed by ARDA in the county where a firm's headquarter is located in a given year, divided by the total population of this county (Cai and Shi, 2019; Callen and Fang, 2015; Chantziaras et al., 2020; Dyreng et al., 2012; Hilary and Hui, 2009; Leventis et al., 2018). ARDA defines total adherents as "all members, including full members, their children and the estimated number of other participants who are not considered members; for example, the 'baptized,' 'those not confirmed,' 'those not eligible for communion,' 'those regularly attending services,' and the like." Our sample firms are mostly multinational companies whose managers, customers or employees come from different states and even other continents. However, social norm theory indicates that managers are strongly influenced by the social norms of the local community where managers live and work, especially religious norms (McGuire et al., 2012). Following Dyreng et al. (2012) and Leventis et al. (2018), we assume that the higher RELIGIOSITY is, the greater the impact of social norms promoted by religions (such as honesty or risk aversion) on firms located in a given county.

While greenwashing data cover the 2005-2019 period, religiosity data are only available for the years 2000 and 2010 due to the frequency of surveys. To get *RELIGIOSTY* values for our sample years prior to 2010, we thus interpolate our religiosity measures using 2000 and 2010 ARDA data to get religiosity estimates for each sample years (Chantziaras et al., 2020; Dyreng et al., 2012; Hilary and Hui, 2009). For years after 2010, we extrapolate our *RELIGIOSITY* estimates following previous research (Chantziaras et al., 2020; Dyreng et al., 2012). This methodology allows us to increase the power of our statistical tests (Hilary and Hui 2009; Leventis et al., 2018). However, we also check that our main results are similar by retaining only the year for which we have a direct measure for religiosity.

## 4. Empirical analysis

### 4.1. Models

In this paper, we examine the impact of religious adherence on greenwashing. We first run a logistic regression model specified as follows:

$$\Pr(GREENWASHING\_BIN_{i,t} = 1) = F\left( \beta_0 + \beta_1 RELIGIOSITY_{i,t} + \sum_k \alpha_k CONTROLS_{i,j}^k + \sum_t \gamma_t Y_i^t + \sum_p \theta_p I_i^p + \varepsilon_{i,t} \right) \quad (1)$$

where $F(x) = \frac{e^x}{(1+e^x)}$ is the cumulative logistic distribution. Our dependent variable is $GREENWASHING\_BIN_{i,t}$, that equals one if the firm *i* in year *t* greenwashes (i.e., if the absolute disclosure ratio is above weighted disclosure ratio) and zero otherwise. Our variable of interest is $RELIGIOSITY_{i,t}$, defined as the number of religious adherents for all denominations in the county where the firm *i* is headquartered in year *t*, divided by the county's total population. $CONTROLS_{i,j}^k$ is a vector of *k* control variables that relate to firm and county characteristics. Specifically, we include *SIZE*, calculated as the natural logarithm of firms' net sales. Large firms are more visible, and are thus more likely to experience external pressures (Kim and Lyon, 2015). In a similar vein and following the previous literature, we include *FOREIGN_SALES*, defined as the percentage of sales to foreign countries (Arouri et al., 2021; Marquis et al., 2016). Firms with a significant proportion of their sales realized abroad are under greater institutional pressure and scrutiny, which can influence environmental publication policy (Marquis et al., 2016). We also control for *ROA*, which is the return on assets as profitability may be linked to greenwashing behaviors (Delmas and Burbano, 2011). Following Marquis et al., (2016) and Arouri et al., (2021), we also control for *CAPITAL_INTENSITY*, calculated as the ratio between net property, plants and equipment and total assets. We control for *ENV_DAMAGES*, measured as total direct environmental costs, scaled by total assets (El Ghoul et al., 2018). Environmental costs data are retrieved from Trucost. Direct environmental costs are defined as environmental costs induced by firms' own operations. They are expressed in US dollars value, and cover six areas of environmental costs: greenhouse gases, waste, water, air pollutants, natural resource use and land and water pollutants. More environmentally damaging firms are less likely to engage in greenwashing as firms with lower environmental performance face greater external scrutiny (Marquis et al., 2016). Finally, we also account for the potential impact of governance on environmental disclosure by including GOVERNANCE, which id Refinitiv's firm governance score.

We also control for demographic variables that are potentially linked with religiosity and that may affect environmental disclosures (Chantziaras et al., 2020). Thus, in order to isolate the impact of religiosity on greenwashing we include *AGE*, *EDUC*, *MALEMIN* and *POP* in all our models. *AGE* is the median age of the county's population in which the firm is headquartered. *POP* is the logarithm of a county's total population, whereas *MALEMIN* is the fraction of male minority population in a county. Finally, *EDUC* is calculated as the fraction of population over 25 years that have completed a bachelor's degree or higher. Lastly, all our models include *p* industry fixed-effects ($I_i^p$) and *t* year fixed-effects ($Y_i^t$), where *p* and *t* are respectively equal to 10 and 14. The literature has already highlighted the significant impact of industrial affiliation on greenwashing behavior (Marquis et al. 2016; Arouri et al. 2021; Pope et al. 2023). We also cluster our standard errors at the firm-level, following previous literature (Adhikari and Agrawal, 2016; Berry-Stölzle and Irlbeck, 2021; Callen and Fang, 2015; Hilary and Hui, 2009).

We then specify the following ordinary least squares regression to evaluate the impact of religious adherence on the magnitude of greenwashing:

$$GREENWASHING\_MAG_{i,t} = \beta_0 + \beta_0 RELIGIOSITY_{i,t} + \sum_k \alpha_k CONTROLS_{i,j}^k + \sum_n \gamma_t Y_{i,j}^n + \sum_p \theta_p I_{i,j}^p + \varepsilon_{i,t} \quad (2)$$

where $GREENWASHING\_MAG_{i,t}$ equals the difference between the absolute disclosure ratio and the weighted disclosure ratio if firm *I's* absolute disclosure ratio is strictly above its weighted disclosure ratio in year *t*, and zero if its absolute disclosure ratio is below its weighted disclosure ratio. We include the same set of control variables ($CONTROLS_{i,j}^{k}$), industry ($I_{i,j}^{p}$) and year ($Y_{i,j}^{n}$) fixed effects as in equation (1) and cluster standard errors at the firm level.

4.2. Summary statistics

We report summary statistics in Table 2. All variables are winsorized at the 1st and 99th percentile to mitigate the potential effect of outliers.

[Insert Table 2 here]

First, it is interesting to note that 10.4% of the companies in our sample appear to greenwash (the mean of *BINARY_GW* is equal to 0.104). In addition, Table 2 shows that sample firms have average net sales of around USD5.4 billion per year. The average *ROA* is 8%. Net property, plants and equipment and foreign sales average around 21% and 31% of total assets, respectively. In Table 3, we report the average religious adherence by state. We observe the highest values of religiosity in Louisiana (67.15), Utah (75.10) and Alabama (77.06), while the lowest levels of religiosity are found in Maine (28.56), Oregon (32.54) and Nevada (35.64), respectively.

[Insert Table 3 here]

Table 4 shows the correlation matrix between all our variables. We see that the correlation coefficients between our control variables are relatively low. We can thus reasonably assume that our subsequent estimates do not suffer from multicollinearity. More importantly, we see that our two proxies of greenwashing (i.e., *GREENWASHING_BIN* and *GREENWASHING_MAG*) are negatively and significantly correlated with *RELIGIOSITY*. This preliminary result seems to provide initial evidence that social norms promoted by religious practices have a negative impact on both the decision to greenwash (-0.124) and the magnitude of greenwashing (-0.125).

[Insert Table 4 here]

4.3. Univariate analysis

In this section, we conduct a univariate analysis to examine whether firms headquartered in counties where religious adherence is high do more greenwashing. We split our total sample according to the degree of religiosity. Specifically, we gather firms located in a county whose religiosity level is located below the median of the distribution in a first group, and the ones for which religiosity level is located above the median in a second group. We then test whether our two variables of interest (*GREENWASHING_BIN* and *GREENWASHING_MAG*) have a significantly different mean between the two previously defined groups. Consistent with our hypothesis, our univariate tests reported in Table 5 seem to indicate that firms headquartered in high religiosity counties are less likely to engage in greenwashing. The magnitude of greenwashing also seems to be lower for these companies. Our univariate results therefore provide a first empirical validation of our hypothesis concerning the link between religiosity and greenwashing. We complement these preliminary results with a multivariate analysis in the next section.

[Insert Table 5 here]

4.4. Multivariate analysis

We first assess the impact of *RELIGIOSITY* on *GREENWASHING_BIN* by estimating the equation (1). Our logistic regression results are reported in Table 6. In column 1, we regress *GREENWASHING_BIN* on *RELIGIOSITY*, without including control variables. In columns 2 and 3, we respectively include firm-level and county-level control variables. Finally, in column 4 we include all our previously defined control variables. The coefficient of *RELIGIOSITY* is negative and significant at the 1% level for all our specifications. Focusing on the fully specified model (Column 4), the interpretation is that for each unit-increase in religious adherence, the estimated odds of a firm engaging in greenwashing decrease by 2.18% ($e^{-0.022} - 1$). These results suggest that firms located in counties where religious adherence is high are less likely to engage in greenwashing. Consistent with the previous literature, we also observe that large firms greenwash less than smaller firms, *ceteris paribus* (Kim and Lyon, 2015; Marquis et al., 2016).

[Insert Table 6 here]

Second, we evaluate the link between *RELIGIOSITY* and *GREENWASHING_MAG* by estimating equation (2). As shown in Table 7, we first estimate the link between *RELIGIOSITY* and *GREENWASHING_MAG* without including control variables (column 1). Then we successively include in the model our firm-level variables (column 2), county-level variables (column 3), and finally all our control variables (column 4). Our preliminary results are confirmed as the coefficient on *RELIGIOSITY* are significantly negative for all estimates. These results corroborate the hypothesis that social norms promoted by religious practice lead to a reduction in greenwashing. It would indeed seem that a stronger religious adherence within the county in which a company is located reduces the magnitude of greenwashing.

This is in line with what we hypothesized in that religious adherence, being associated with the social norms of honesty and risk aversion, should be negatively linked with manipulative and risky behaviors.

[Insert Table 7 here]

4.5. Robustness checks

Our results pinpoint a clear relationship between religious adherence and greenwashing. We perform several additional tests to check the robustness of our results. In Table 8, we first estimate in columns 1 and 2 equations (1) and (2) only by using the year 2010, which is the only sample year for which we directly have survey data on religiosity. Following previous literature (Dyreng et al., 2012; Hilary and Hui, 2009), we thus aim to evaluate whether our results are impacted by our approximation of religiosity by interpolation and extrapolation. Second, we seek to examine whether our results could be explained by possible reverse causality by implementing two tests inspired by the work of Cai and Shi (2019). Indeed, if we postulate that firms located in counties with a high degree of religiosity adopt more conservative decisions, it is also possible to argue that firms with more conservative management may tend to locate more easily in counties with higher religious adherence. Firms belonging to the mining, agriculture or manufacturing sectors are more likely to choose their location because of considerations related to production constraints, especially regarding inputs and distribution (John et al., 2011). As a result, we assume that these firms are less likely to make location decisions based on religiosity (Cai and Shi 2019). In columns 3 and 4, we thus estimate equations (1) and (2) by restricting our full sample to firms operating in these sectors. In a similar vein, in columns 5 and 6, we limit our sample to firms with total assets above USD 1 billion. These larger companies are also less likely to choose their location based on religiosity, as the costs of relocation are significant (John et al., 2011; Cai and Shi, 2019). Results featured in Table 8 highlight the robustness of our previous findings as the coefficient on *RELIGIOSITY* is always significantly negative at the 1% level, except for one specification.

To further reduce potential endogeneity concerns, we reproduce our main regression results using a two-stage least squares (2SLS) approach. Following Hilary and Hui (2009) and Jiang et al. (2018), we use as instrumental variables our *RELIGIOSITY* measure lagged by three years and the county population (*POP*) lagged by three years. To further support our choice of instruments, we perform the following two tests: (1) a Kleibergen-Paap LM (2006) instrument relevance test to ensure the relevance of our instruments (i.e., high correlations between the instruments and adjusted *RELIGIOSITY*), and (2) a Hansen *J*-test to check for the orthogonality condition (i.e., no significant correlation between the instruments and the residuals in the stage 2 regression). The relevance of our instruments is confirmed as indicated by the p-value ($p$=0.000) of the Kleibergen-Paap LM test. The Hansen J-test fails to reject the orthogonality condition (p=0.819), suggesting our instruments are valid. We report our results in columns 7 (Stage 2) and 8 (Stage 1) of Table 8. Our previous conclusions are confirmed as evidenced by the negative and statistically significant coefficient on *RELIGIOSITY*.

[Insert Table 8 here]

4.6. Transmission channels analysis

At this stage, while we identified an impact of religious adherence on greenwashing, it is still impossible to determine specifically which social norm promoted by religious adherence explains this relationship. In order to distinguish between the effect of honesty and risk aversion on greenwashing, we thus run complementary tests by using different dependent variables, following Dyreng et al., (2012).

If the promotion of more ethical standards prevails in explaining the negative impact of religiosity on greenwashing, we expect that firms located in high religious adherence counties will be more inclined to minimize the absolute difference between the absolute disclosure ratio and the weighted disclosure ratio, i.e., to minimize the difference between these ratios whatever the sign of this difference in order to depict the most accurate representation of reality. It is possible to make an analogy with the earnings management literature. Previous research has posited that if religious adherence influences earnings management through the honesty channel, religiosity should have a negative impact on both income-increasing and income-decreasing earnings management (Cai et al., 2019; Dyreng et al., 2012; McGuire et al., 2012). Our two proxies for greenwashing (*GREENWASHING_BIN* and *GREENWASHING_MAG*) are built on the principle that a greenwashing situation is observed only when the absolute disclosure ratio is higher than the weighted disclosure ratio. But in some cases, the absolute disclosure ratio is lower than the weighted disclosure ratio, which corresponds to a situation where the firm understates its environmental performance by disclosing proportionally more of it more harmful indicators with respect to its overall environmental impact. In this situation of "brownwashing" (Kim and Lyon, 2015), environmental disclosure also does not fully reflect the reality of its environmental impact, as in the case of greenwashing. If the honesty channel prevails, we should observe a negative impact of religiosity on both greenwashing and brownwashing as the anti-manipulative ethos promoted by religions should lead managers to disclose environmental information that gives the most accurate picture of their environmental impact. In order to check whether the impact of religiosity on greenwashing can be explained by the honesty channel, we thus specify the following model:

$$ABS\_SD\_GW_{i,t} = \beta_0 + \beta_0 RELIGIOSITY_{i,t} + \sum_k CONTROLS_{i,j}^k + \sum_n \alpha_k \gamma_t Y_{i,j}^n + \sum_p \theta_p I_{i,j}^p + \varepsilon_{i,t} \quad (3)$$

where $ABS\_SD\_GW_{i,t}$ is a firm's absolute selective disclosure, computed as the absolute difference absolute and weighted disclosure ratio. $CONTROLS_{i,j}^k$ is the same set of control variables as defined previously. We also include industry and year fixed effects and cluster standard errors at the firm level. If religious adherence influences greenwashing through the honesty channel, *RELIGIOSITY* should have a positive impact on *ABS_SD_GW*. We report estimation results in Table 9. We first regress *ABS_SD_GW* on *RELIGIOSITY* only in column 1. We then include our firm-level variables in column 2, our county-level variables in column 3, and all our control variables in column 4.

[Insert Table 9 here]

We find that the coefficient on *RELIGIOSITY*, although negative, is not statistically significant for any of our model specifications. This tends to invalidate the explanation that greenwashing is impacted by religiosity because of the promotion of more conservative moral standards. Rather, this result would indicate that religiosity impacts greenwashing behaviors through risk aversion.

In a similar way, we estimate multinomial logistic regressions to determine which type of selective disclosure (brownwashing or greenwashing) is influenced by religiosity. Our model is as follows:

$$\Pr(SD\_CHOICE_{i,t}) = F\left(\beta_0 + \beta_1 RELIGIOSITY_{i,t} + \sum_k \alpha_k CONTROLS_{i,j}^k + \sum_t \gamma_t Y_i^t + \sum_p \theta_p I_i^p + \varepsilon_{i,t}\right) \quad (4)$$

where SD_CHOICE is a trinary variable which equals 0 if the firm brownwashes, i.e., if the absolute disclosure ratio is inferior to the weighted disclosure ratio. The variable equals 1 if the firm is not engaged in selective disclosure, i.e., if the absolute disclosure ratio is equal to the weighted disclosure ratio. Finally, the variable equals 2 if the firm is engaged in greenwashing, i.e., if the absolute disclosure ratio is superior to the weighted disclosure ratio. The baseline category is the case where *SD_CHOICE* = 1. This model allows us to determine whether *RELIOGISITY* has an impact on both positive and negative selective disclosure. Our results are reported in Table 10.

[Insert Table 10 here]

Consistent with our previous results, *RELIGIOSITY* only has a significant impact on selective disclosure for the case where *SD_CHOICE* = 2, i.e., for greenwashing. Again, this result is in line with the risk aversion channel, since it appears that religious adherence has an impact on greenwashing but not on brownwashing.

To further investigate the relative importance of risk aversion and honesty in explaining the relationship between religiosity and greenwashing, we exploit the religious diversity of our sample of US firms. Indeed, The US is largely Christian, but the Christian population is quite heterogeneous, separated into Catholics and Protestants (Hilary and Hui, 2009). It has been widely documented that Protestant populations appear to have a higher level of risk aversion than Catholic populations, leading to lower risk-taking (Shu et al., 2012), more generous dividend policies (Ucar, 2016), less lottery-type stock holdings (Kumar et al., 2011), lower investment rates (Hilary and Hui, 2009) or less earnings management (Cai et al., 2019). If we assume that religious adherence has a homogenous impact on the promotion of an anti-manipulative ethos, we can break down religious adherence into Catholics and Protestants to isolate the role played by risk aversion in the relationship between religiosity and our dependent variables (Hilary and Hui, 2009; Dyreng et al., 2012). We thus estimate equation (1) and (2) by replacing *RELIGIOSITY* by *CATHRT* (columns 1 and 3) and *MAINRT* (column 2 and 4), which are the ratio between Catholic (*CATHRT)* and Protestant (*MAINRT)* adherents in the county where the firm's headquarter is located in a given year, divided by the county's total population. If the risk aversion channel explains the negative impact of religious adherence on greenwashing, we should observe a higher coefficient for MAINRT than for CATHRT. Results are provided in Table 11.

[Insert Table 11 here]

The dependent variable is *GREENWASHING_BIN* in columns 1 and 2 and *GREENWASHING_MAG* in columns 3 and 4. We find that *MAINRT* has a negative and significant effect on both *GREENWASHING_BIN* (at the 1% level) and *GREENWASHING_MAG* (at the 5% level), whereas the impact of CATHRT is not statistically different from zero. In addition, we test whether coefficients for *MAINRT* and *CATHRT* are significantly different. The coefficient for *MAINRT* seems to be significantly lower than for CATHRT only for regressions that have *GREENWASHING_BIN* as dependent variables. Taken together, these results suggest that Protestant religious adherence has a more negative impact on greenwashing behavior than Catholic religious adherence. This highlights the role played by risk aversion in explaining the impact of religiosity on greenwashing, even if the results concerning *GREENWASHING_MAG* do not seem significant. This can be explained by the fact that the decision to engage in greenwashing is a clearer decision than the modulation of the level of greenwashing.

## 5. Conclusion

5.1. Findings

In this paper, we investigate how religious adherence affects corporate greenwashing behaviors. Using a sample of US firms between 2005 and 2019, we find that firms located in counties where religious adherence is high are less likely to engage in greenwashing and that a stronger religious adherence within the county in which a company is located reduces the magnitude of greenwashing. We also show that the impact of religiosity on greenwashing materializes through the channel of risk aversion.

5.2. Implications for theory and practice

Our results have important implications for regulators and market participants. At a time of increased public interest in environmental issues, there is a risk that firms may choose to engage in strategic disclosure and communication in order to appear 'greener' than they really are (Arouri et al., 2021). Such behavior poses various threats to the functioning of markets. Indeed, the fear of greenwashing could produce negative effects on consumers and investors' confidence in "green" activities and products. In addition, biased disclosure of environmental performance could negatively influence market efficiency by distorting the link between market prices and company fundamentals. Such a distortion would then lead to a misallocation of financial resources and negatively impact general levels of welfare. Because of these risks, understanding the drivers and motivations underlying greenwashing behavior by firms is necessary, and our work contributes to this understanding.

On theoretical grounds, our findings align with social norm theory (Dyreng et al., 2012; Kohlberg, 1984; Sunstein, 1996; McGuire et al., 2012) and neo-institutionalist theory (Meyer and Rowan 1977; DiMaggio and Powell 1983) in that they show that religiosity shapes firms' propensity to engage in greenwashing behavior. In doing so, we contribute to this body of knowledge by evidencing that religion is a powerful social norm.

As far as policymaking is concerned, understanding how religious social norms affect managers may help regulators, contract designers, and financial market participants to shape appropriate standards (Sunder, 2005). Indeed, knowing how religious adherence might impact the propensity to engage in manipulative behavior may help policymakers anticipate the potential consequences of proposed legislation and thus evaluate their soundness.

Our findings also have practical implications for investors. Indeed, investors who care about extra-financial performance may decide to use religiosity as an additional criterion in their investment screening process to assess the risk of greenwashing. In a context in which firms communicate heavily on corporate social responsibility, being able to assess the quality of such actions as well as the veracity of some claims is key for investors, whether or not they have CSR mandates. Indeed, CSR controversies and the exposure of greenwashing may lead to negative cumulative abnormal returns (Du, 2015). As such, having a knowledge of how religious adherence may influence the likelihood of such behavior is relevant.

5.3. Mains limits and avenues for future research

Our study has some limitations that constitute avenues for further research. First, we restrict our analysis to the US context. While this has the advantage of reducing potential heterogeneity issues, it could be interesting to use cross-country data to check how these findings hold internationally. Second, our study is quantitative in nature. As such, it could be complemented by qualitative studies such as interviews or experiments. Third, our sample is limited by the availability of Trucost data. As such, it would be interesting to develop new measures of greenwashing so as to expand the analysis to a broader sample. Finally, religion could impact other manipulative non-financial disclosures unrelated to environmental issues. As such, it would be interesting to study whether religiosity impacts other unethical practices such as social washing (Troje and Gluch, 2020).

Table 1. Sample description

| State | N | Industry | N | Year | N |
|---|---|---|---|---|---|
| Alabama | 22 | Basic materials | 337 | 2005 | 282 |
| Arkansas | 54 | Consumer discretionary | 1,344 | 2006 | 286 |
| Arizona | 106 | Consumer staples | 425 | 2007 | 315 |
| California | 881 | Energy | 201 | 2008 | 371 |
| Colorado | 159 | Financials | 891 | 2009 | 421 |
| Connecticut | 176 | Health care | 698 | 2010 | 459 |
| DC | 32 | Industrials | 1,409 | 2011 | 482 |
| Delaware | 18 | Real estate | 139 | 2012 | 467 |
| Florida | 268 | Technology | 812 | 2013 | 472 |
| Georgia | 197 | Telecommunications | 184 | 2014 | 488 |
| Iowa | 28 | Utilities | 110 | 2015 | 502 |
| Idaho | 9 | | | 2016 | 507 |
| Illinois | 536 | | | 2017 | 515 |
| Indiana | 99 | | | 2018 | 504 |
| Kansas | 28 | | | 2019 | 479 |
| Kentucky | 44 | | | | |
| Louisiana | 26 | | | | |
| Massachusetts | 286 | | | | |
| Maryland | 87 | | | | |
| Maine | 8 | | | | |
| Michigan | 123 | | | | |
| Minnesota | 199 | | | | |
| Missouri | 121 | | | | |
| North Carolina | 150 | | | | |
| Nebraska | 45 | | | | |
| New Jersey | 166 | | | | |
| Nevada | 23 | | | | |
| New York | 610 | | | | |
| Ohio | 325 | | | | |
| Oklahoma | 12 | | | | |
| Oregon | 24 | | | | |
| Pennsylvania | 382 | | | | |
| Rhode Island | 57 | | | | |
| South Carolina | 25 | | | | |
| Tennessee | 106 | | | | |
| Texas | 594 | | | | |
| Utah | 30 | | | | |
| Virginia | 207 | | | | |
| Washington | 142 | | | | |
| Wisconsin | 144 | | | | |
| Total | 6,550 | Total | 6,550 | Total | 6,550 |

Table 1 presents our sample breakdown by state, industry, and year.

Table 2. Descriptive statistics

| VARIABLES | N | Mean | Q1 | Median | Q3 | Max | Min |
|---|---|---|---|---|---|---|---|
| GREENWASHING_BIN | 6,550 | 0.119 | 0.000 | 0.000 | 0.000 | 1.000 | 0.000 |
| GREENWASHING_MAG | 6,550 | 3.010 | 0.000 | 0.000 | 0.000 | 66.000 | 0.000 |
| RELIGIOSITY | 6,550 | 50.652 | 43.587 | 51.121 | 59.422 | 77.828 | 23.270 |
| CAPITAL_INTENSITY | 6,550 | 20.907 | 5.893 | 14.287 | 30.780 | 84.735 | 0.250 |
| SIZE | 6,550 | 15.721 | 14.820 | 15.594 | 16.496 | 18.902 | 12.763 |
| ROA | 6,550 | 7.989 | 3.960 | 7.235 | 11.310 | 30.480 | -11.160 |
| FOREIGN_SALES | 6,550 | 31.550 | 0.000 | 29.405 | 51.630 | 99.030 | 0.000 |
| ENV_DAMAGES | 6,550 | 0.540 | 0.019 | 0.089 | 0.253 | 9.378 | 0.000 |
| GOVERNANCE | 6,550 | 52.443 | 35.370 | 53.500 | 70.470 | 98.610 | 1.240 |
| AGE | 6,550 | 36.796 | 34.700 | 36.700 | 38.400 | 45.000 | 31.500 |
| EDUC | 6,550 | 30.174 | 0.590 | 33.200 | 46.100 | 61.900 | 0.203 |
| MALEMIN | 6,550 | 27.807 | 18.899 | 28.155 | 34.056 | 55.423 | 5.040 |
| POP | 6,550 | 13.899 | 13.435 | 13.869 | 14.410 | 16.103 | 11.547 |

Table 2 presents descriptive statistics for our sample variables. GREENWASHING_BIN is a binary variable that equals one if a firm greenwashes and zero otherwise. GREENWASHING_MAG is the difference between the absolute disclosure ratio and the weighted disclosure ratio if a firm's absolute disclosure ratio is strictly above its weighted disclosure ratio, and zero otherwise. RELIGIOSITY is the number of religious adherents for all denominations in the county where a firm is headquartered divided by the county's total population. CAPITAL_INTENSITY, is the ratio between net property, plants and equipment and total assets. SIZE is the natural logarithm of firms' net sales. ROA is the return on assets. FOREIGN_SALES is the percentage of sales to foreign countries. ENV_DAMAGES represents total direct environmental costs scaled by total assets. GOVERNANCE is Refinitiv's firm governance score. AGE is the median age of the county's population in which a firm is headquartered. EDUC is calculated as the fraction of population over 25 years that have completed a bachelor's degree or higher. MALEMIN is the fraction of male minority population in a county. POP is the logarithm of a county's total population.

Table 3. County Religiosity Descriptive statistics

| | Reliogisity | | |
|---|---|---|---|
| Alabama | 77.061 | Michigan | 46.526 |
| Arkansas | 53.853 | Minnesota | 52.341 |
| Arizona | 38.937 | Missouri | 50.820 |
| California | 43.665 | North Carolina | 50.197 |
| Colorado | 40.058 | Nebraska | 52.858 |
| Connecticut | 57.267 | New Jersey | 58.857 |
| DC | 50.350 | Nevada | 35.644 |
| Delaware | 46.823 | New York | 44.337 |
| Florida | 40.473 | Ohio | 47.593 |
| Georgia | 65.895 | Oklahoma | 63.337 |
| Iowa | 44.421 | Oregon | 32.543 |
| Idaho | 42.674 | Pennsylvania | 55.645 |
| Illinois | 57.568 | Rhode Island | 52.660 |
| Indiana | 43.745 | South Carolina | 48.365 |
| Kansas | 54.455 | Tennessee | 62.844 |
| Kentucky | 54.620 | Texas | 59.425 |
| Louisiana | 67.148 | Utah | 75.098 |
| Massachusetts | 58.946 | Virginia | 48.643 |
| Maryland | 42.255 | Washington | 37.657 |
| Maine | 28.558 | Wisconsin | 49.925 |

Table 3 presents descriptive statistics for religious adherence by state, measured as the number of religious adherents for all denominations in the county where a firm is headquartered divided by the county's total population.

Table 4. Correlations

| | (1) | (2) | (3) | (4) | (5) | (6) | (7) | (8) | (9) | (10) | (11) | (12) | (13) |
|---|---|---|---|---|---|---|---|---|---|---|---|---|---|
| (1) GREENWASHING_BIN | 1 | | | | | | | | | | | | |
| (2) GREENWASHING_MAG | 0.749*** | 1 | | | | | | | | | | | |
| (3) RELIGIOSITY | -0.131*** | -0.130*** | 1 | | | | | | | | | | |
| (4) FOREIGN_SALES | 0.010 | -0.015 | -0.114*** | 1 | | | | | | | | | |
| (5) SIZE | 0.206*** | 0.148*** | -0.003 | -0.020 | 1 | | | | | | | | |
| (6) ROA | -0.005 | 0.012 | -0.012 | 0.137*** | -0.035*** | 1 | | | | | | | |
| (7) CAPITAL_INTENSITY | -0.087*** | -0.066*** | 0.087*** | -0.075*** | 0.084*** | 0.054*** | 1 | | | | | | |
| (8) ENV_DAMAGES | -0.099*** | -0.085*** | 0.103*** | 0.003 | 0.024** | -0.008 | 0.441*** | 1 | | | | | |
| (9) GOVERNANCE | 0.032*** | 0.104*** | -0.016 | 0.078*** | 0.317*** | 0.015 | 0.088*** | 0.099*** | 1 | | | | |
| (10) AGE | 0.124*** | 0.038*** | -0.180*** | 0.030** | -0.044*** | 0.031** | -0.154*** | -0.096*** | 0.004 | 1 | | | |
| (11) POP | 0.059*** | 0.061*** | 0.075*** | 0.095*** | 0.032** | -0.013 | -0.011 | -0.000 | 0.051*** | -0.282*** | 1 | | |
| (12) MALEMIN | 0.085*** | 0.062*** | 0.033*** | 0.021* | 0.098*** | 0.004 | -0.058*** | -0.018 | 0.056*** | -0.324*** | 0.296*** | 1 | |
| (13) EDUC | 0.183*** | 0.181*** | -0.150*** | 0.053*** | 0.015 | 0.004 | -0.106*** | -0.085*** | 0.137*** | 0.193*** | 0.060*** | 0.204*** | 1 |

Table 4 presents pairwise correlations between our sample variables. GREENWASHING_BIN is a binary variable that equals one if a firm greenwashes and zero otherwise. GREENWASHING_MAG is the difference between the absolute disclosure ratio and the weighted disclosure ratio if a firm's absolute disclosure ratio is strictly above its weighted disclosure ratio, and zero otherwise. RELIGIOSITY is the number of religious adherents for all denominations in the county where a firm is headquartered divided by the county's total population. CAPITAL_INTENSITY, is the ratio between net property, plants and equipment and total assets. SIZE is the natural logarithm of firms' net sales. ROA is the return on assets. FOREIGN_SALES is the percentage of sales to foreign countries. ENV_DAMAGES represents total direct environmental costs scaled by total assets. GOVERNANCE is Refinitiv's firm governance score. AGE is the median age of the county's population in which a firm is headquartered. EDUC is calculated as the fraction of population over 25 years that have completed a bachelor's degree or higher. MALEMIN is the fraction of male minority population in a county. POP is the logarithm of a county's total population.

Table 5. Univariate analysis

|  | N_1 | N_2 | Mu_1 | Mu_2 | Mu_1 – Mu_2 | p |
|---|---|---|---|---|---|---|
| GREENWASHING_BIN | 3,368 | 3,182 | 0.159 | 0.077 | 0.082 | 0.000 |
| GREENWASHING_MAG | 3,368 | 3,182 | 4.116 | 1.840 | 2.277 | 0.000 |
| FOREIGN_SALES | 3,368 | 3,182 | 35.410 | 27.465 | 7.945 | 0.000 |
| SIZE | 3,368 | 3,182 | 15.698 | 15.746 | -0.048 | 0.126 |
| ROA | 3,368 | 3,182 | 8.150 | 7.818 | 0.333 | 0.038 |
| CAPITAL_INTENSITY | 3,368 | 3,182 | 19.289 | 22.620 | -3.332 | 0.000 |
| ENV_DAMAGES | 3,368 | 3,182 | 0.418 | 0.668 | -0.250 | 0.000 |
| GOVERNANCE | 3,368 | 3,182 | 52.872 | 51.990 | 0.882 | 0.107 |
| AGE | 3,368 | 3,182 | 37.195 | 36.373 | 0.822 | 0.000 |
| POP | 3,368 | 3,182 | 13.789 | 14.014 | -0.225 | 0.000 |
| MALEMIN | 3,368 | 3,182 | 28.374 | 27.207 | 1.168 | 0.000 |
| EDUC | 3,368 | 3,182 | 32.459 | 27.756 | 4.703 | 0.000 |
| N | 6,550 | | | | | |

Table 5 presents results from a univariate analysis differentiating between firms located in a county whose religiosity level is located below the median of the distribution in a first group (Mu_1), and the ones for which religiosity level is located above the median in a second group (Mu_2). GREENWASHING_BIN is a binary variable that equals one if a firm greenwashes and zero otherwise. GREENWASHING_MAG is the difference between the absolute disclosure ratio and the weighted disclosure ratio if a firm's absolute disclosure ratio is strictly above its weighted disclosure ratio, and zero otherwise. RELIGIOSITY is the number of religious adherents for all denominations in the county where a firm is headquartered divided by the county's total population. CAPITAL_INTENSITY, is the ratio between net property, plants and equipment and total assets. SIZE is the natural logarithm of firms' net sales. ROA is the return on assets. FOREIGN_SALES is the percentage of sales to foreign countries. ENV_DAMAGES represents total direct environmental costs scaled by total assets. GOVERNANCE is Refinitiv's firm governance score. AGE is the median age of the county's population in which a firm is headquartered. EDUC is calculated as the fraction of population over 25 years that have completed a bachelor's degree or higher. MALEMIN is the fraction of male minority population in a county. POP is the logarithm of a county's total population.

Table 6. Religious adherence and greenwashing likelihood

|  | (1) GREENWASHING_BIN | (2) GREENWASHING_BIN | (3) GREENWASHING_BIN | (4) GREENWASHING_BIN |
|---|---|---|---|---|
| RELIGIOSITY | -0.026*** | -0.025*** | -0.022*** | -0.022*** |
|  | (-3.28) | (-3.07) | (-2.88) | (-2.64) |
| FOREIGN_SALES |  | -0.001 |  | -0.002 |
|  |  | (-0.24) |  | (-0.49) |
| SIZE |  | 0.543*** |  | 0.528*** |
|  |  | (6.97) |  | (6.78) |
| ROA |  | 0.014 |  | 0.013 |
|  |  | (0.99) |  | (0.92) |
| CAPITAL_INTENSITY |  | -0.014** |  | -0.013 |
|  |  | (-2.35) |  | (-2.09) |
| ENV_DAMAGES |  | -0.262 |  | -0.257 |
|  |  | (-1.02) |  | (-1.03) |
| GOVERNANCE |  | 0.010** |  | 0.010** |
|  |  | (2.30) |  | (2.23) |
| AGE |  |  | -0.021 | -0.017 |
|  |  |  | (-0.53) | (-0.44) |
| POP |  |  | 0.121 | 0.095 |
|  |  |  | (0.99) | (0.73) |
| MALEMIN |  |  | 0.004 | -0.003 |
|  |  |  | (0.42) | (-0.34) |
| EDUC |  |  | 0.022*** | 0.017* |
|  |  |  | (2.41) | (1.92) |
| Constant | -3.472*** | -11.510*** | -4.584* | -11.990*** |
|  | (-3.99) | (-7.85) | (-1.83) | (-4.41) |
| Industry FE | yes | yes | yes | yes |
| Year FE | yes | yes | yes | yes |
| N | 6,550 | 6,550 | 6,550 | 6,550 |
| pseudo $R^2$ | 0.094 | 0.172 | 0.104 | 0.177 |

Table 6 presents results from our multivariate analysis using logistic regression to assess the impact of religious adherence on greenwashing likelihood. GREENWASHING_BIN is a binary variable that equals one if a firm greenwashes and zero otherwise. RELIGIOSITY is the number of religious adherents for all denominations in the county where a firm is headquartered divided by the county's total population. CAPITAL_INTENSITY, is the ratio between net property, plants and equipment and total assets. SIZE is the natural logarithm of firms' net sales. ROA is the return on assets. FOREIGN_SALES is the percentage of sales to foreign countries. ENV_DAMAGES represents total direct environmental costs scaled by total assets. GOVERNANCE is Refinitiv's firm governance score. AGE is the median age of the county's population in which a firm is headquartered. EDUC is calculated as the fraction of population over 25 years that have completed a bachelor's degree or higher. MALEMIN is the fraction of male minority population in a county. POP is the logarithm of a county's total population. T-statistics are in parentheses. *,**, and *** represent statistical significance at the 10%, 5%, and 1% level, respectively.

Table 7. Religious adherence and greenwashing magnitude

|  | (1) GREENWASHING_MAG | (2) GREENWASHING_MAG | (3) GREENWASHING_MAG | (4) GREENWASHING_MAG |
|---|---|---|---|---|
| RELIGIOSITY | -0.099*** | -0.097*** | -0.087*** | -0.086*** |
|  | (-3.22) | (-3.21) | (-2.85) | (-2.86) |
| FOREIGN_SALES |  | -0.021 |  | -0.026* |
|  |  | (-1.39) |  | (-1.78) |
| SIZE |  | 1.293*** |  | 1.247*** |
|  |  | (4.68) |  | (4.50) |
| ROA |  | 0.055 |  | 0.045 |
|  |  | (1.16) |  | (0.96) |
| CAPITAL_INTENSITY |  | -0.032* |  | -0.025 |
|  |  | (-1.94) |  | (-1.47) |
| ENV_DAMAGES |  | -0.187 |  | -0.185 |
|  |  | (-1.47) |  | (-1.40) |
| GOVERNANCE |  | 0.023** |  | 0.023** |
|  |  | (2.02) |  | (2.03) |
| AGE |  |  | -0.055 | -0.063 |
|  |  |  | (-0.55) | (-0.65) |
| POP |  |  | 0.524** | 0.510** |
|  |  |  | (1.98) | (1.99) |
| MALEMIN |  |  | -0.011 | -0.027 |
|  |  |  | (-0.39) | (-0.98) |
| EDUC |  |  | 0.139*** | 0.136*** |
|  |  |  | (3.34) | (3.28) |
| Constant | 3.077* | -15.930*** | -2.163 | -19.470*** |
|  | (1.89) | (-3.54) | (-0.38) | (-2.87) |
| Industry FE | yes | yes | yes | yes |
| Year FE | yes | yes | yes | yes |
| N | 6,550 | 6,550 | 6,550 | 6,550 |
| adj. $R^2$ | 0.048 | 0.077 | 0.062 | 0.090 |

Table 7 presents results from our multivariate analysis to assess the impact of religious adherence on greenwashing magnitude. *GREENWASHING_MAG* is the difference between the absolute disclosure ratio and the weighted disclosure ratio if a firm's absolute disclosure ratio is strictly above its weighted disclosure ratio, and zero otherwise. *RELIGIOSITY* is the number of religious adherents for all denominations in the county where a firm is headquartered divided by the county's total population. *CAPITAL_INTENSITY*, is the ratio between net property, plants and equipment and total assets. *SIZE* is the natural logarithm of firms' net sales. *ROA* is the return on assets. *FOREIGN_SALES* is the percentage of sales to foreign countries. *ENV_DAMAGES* represents total direct environmental costs scaled by total assets. *GOVERNANCE* is Refinitiv's firm governance score. *AGE* is the median age of the county's population in which a firm is headquartered. *EDUC* is calculated as the fraction of population over 25 years that have completed a bachelor's degree or higher. *MALEMIN* is the fraction of male minority population in a county. *POP* is the logarithm of a county's total population. T-statistics are in parentheses. *,**, and *** represent statistical significance at the 10%, 5%, and 1% level, respectively.

Table 8. Robustness tests

| | (1) GREENWASHING_BIN 2010 only | (2) GREENWASHING_MAG 2010 only | (3) GREENWASHING_BIN mining, agriculture, manufacturing | (4) GREENWASHING_MAG mining, agriculture, manufacturing | (5) GREENWASHING_BIN >USD1B | (6) GREENWASHING_MAG >USD1B | (7) GREENWASHING_MAG 2SLS – Stage 2 | (8) 2SLS – Stage 1 |
|---|---|---|---|---|---|---|---|---|
| RELIGIOSITY | -0.062*** | -0.049* | -0.053*** | -0.041** | -0.022*** | -0.089*** | -0.097** | |
| | (-2.72) | (-1.67) | (-3.69) | (-2.49) | (-2.66) | (-2.87) | (-2.53) | |
| FOREIGN_SALES | -0.025*** | -0.046* | 0.002 | -0.003 | -0.002 | -0.027* | -0.036* | -0.001 |
| | (-2.67) | (-1.95) | (0.38) | (-0.39) | (-0.53) | (-1.76) | (-1.91) | (-0.34) |
| SIZE | 0.659*** | 1.115*** | 0.088 | 0.183 | 0.508*** | 1.206*** | 1.327*** | -0.079 |
| | (4.33) | (3.24) | (0.58) | (0.81) | (6.35) | (4.05) | (3.86) | (-1.15) |
| ROA | 0.011 | 0.006 | 0.005 | 0.006 | 0.014 | 0.049 | 0.044 | 0.010 |
| | (0.27) | (0.14) | (0.24) | (0.22) | (1.01) | (0.96) | (0.70) | (0.93) |
| CAPITAL_INTENSITY | -0.042*** | -0.046** | -0.030 | -0.026** | -0.013** | -0.025 | -0.025 | 0.008 |
| | (-2.77) | (-2.30) | (-1.58) | (-1.99) | (-2.11) | (-1.45) | (-1.05) | (1.63) |
| ENV_DAMAGES | -0.041 | 0.031 | 0.160 | 0.142** | -0.251 | -0.191 | -0.232 | 0.074 |
| | (-0.11) | (0.29) | (1.39) | (2.05) | (-1.02) | (-1.43) | (-1.28) | (1.24) |
| GOVERNANCE | 0.016 | -0.004 | 0.016** | 0.009 | 0.010** | 0.024** | 0.030** | -0.002 |
| | (1.44) | (-0.23) | (2.05) | (1.42) | (2.24) | (2.04) | (2.02) | (-0.70) |
| AGE | -0.051 | -0.062 | 0.084 | 0.071 | -0.020 | -0.066 | -0.052 | -0.133*** |
| | (-0.74) | (-0.70) | (1.54) | (1.54) | (-0.50) | (-0.67) | (-0.43) | (-3.96) |
| POP | -0.035 | -0.165 | 0.153 | 0.181 | 0.091 | 0.502* | 0.594** | 23.250*** |
| | (-0.12) | (-0.56) | (0.62) | (0.66) | (0.70) | (1.93) | (1.98) | (8.30) |
| MALEMIN | -0.038 | -0.032 | 0.014 | 0.011 | -0.004 | -0.028 | -0.030 | 0.052*** |
| | (-1.32) | (-1.10) | (1.03) | (0.87) | (-0.41) | (-0.99) | (-0.86) | (5.98) |
| EDUC | 0.026 | 0.066* | -0.002 | 0.018 | 0.018* | 0.140*** | 0.144*** | -0.110*** |
| | (1.56) | (1.74) | (-0.10) | (0.79) | (1.96) | (3.24) | (3.17) | (-10.96) |
| RELIGIOSITY (lagged 3Y) | | | | | | | | 1.064*** |
| | | | | | | | | (139.29) |
| POP (lagged 3Y) | | | | | | | | -23.550*** |
| | | | | | | | | (-8.41) |
| Constant | -6.327 | -6.730 | -8.284* | -5.768 | -11.520*** | -18.510*** | -21.310** | 3.292 |
| | (-1.02) | (-0.87) | (-1.83) | (-1.19) | (-4.19) | (-2.59) | (-2.47) | (1.40) |
| Industry FE | yes | yes | yes | yes | yes | yes | yes | yes |
| Year FE | no | no | yes | yes | yes | yes | yes | yes |
| N | 433 | 459 | 3,150 | 3,171 | 6,274 | 6,384 | 5,008 | 5,008 |
| adj. $R^2$ | | 0.053 | | 0.060 | | 0.089 | 0.089 | |
| pseudo $R^2$ | 0.259 | | 0.176 | | 0.172 | | | |
| Hansen J-test p-value | | | | | | | | 0.000 |
| Kleibergen-Papp LM p-value | | | | | | | | 0.819 |

Table 8 presents various robustness tests. In columns 1 and 2, we first estimate equations (1) and (2) only by using the year 2010. In columns 3 and 4, we estimate equations (1) and (2) by restricting our full sample to firms operating in the mining, agriculture, and manufacturing sectors. In columns 5 and 6, we limit our sample to firms with total assets above USD 1 billion. GREENWASHING_BIN is a binary variable that equals one if a firm greenwashes and zero otherwise. GREENWASHING_MAG is the difference between the absolute disclosure ratio and the weighted disclosure ratio if a firm's absolute disclosure ratio is strictly above its weighted disclosure ratio, and zero otherwise. RELIGIOSITY is the number of religious adherents for all denominations in the county where a firm is headquartered divided by the county's total population. CAPITAL_INTENSITY, is the ratio between net property, plants and equipment and total assets. SIZE is the natural logarithm of firms' net sales. ROA is the return on assets. FOREIGN_SALES is the percentage of sales to foreign countries. ENV_DAMAGES represents total direct environmental costs scaled by total assets. GOVERNANCE is Refinitiv's firm governance score. AGE is the median age of the county's population in which a firm is headquartered. EDUC is calculated as the fraction of population over 25 years that have completed a bachelor's degree or higher. MALEMIN is the fraction of male minority population in a county. POP is the logarithm of a county's total population. T-statistics are in parentheses. *,**, and *** represent statistical significance at the 10%, 5%, and 1% level, respectively.

Table 9. Transmission channels analysis

| | (1) ABS_SD_GW | (2) ABS_SD_GW | (3) ABS_SD_GW | (4) ABS_SD_GW |
|---|---|---|---|---|
| RELIGIOSITY | -0.089 | -0.060 | -0.090 | -0.050 |
| | (-1.45) | (-1.05) | (-1.34) | (-0.83) |
| FOREIGN_SALES | | 0.188*** | | 0.190*** |
| | | (6.94) | | (6.89) |
| SIZE | | 4.296*** | | 4.269*** |
| | | (8.71) | | (8.57) |
| ROA | | 0.008 | | 0.003 |
| | | (0.11) | | (0.04) |
| CAPITAL_INTENSITY | | 0.050 | | 0.051 |
| | | (1.45) | | (1.49) |
| ENV_DAMAGES | | 2.469*** | | 2.427*** |
| | | (3.32) | | (3.33) |
| GOVERNANCE | | 0.160*** | | 0.161*** |
| | | (6.86) | | (6.97) |
| AGE | | | 0.120 | 0.125 |
| | | | (0.39) | (0.47) |
| POP | | | -0.344 | -0.756 |
| | | | (-0.42) | (-1.07) |
| MALEMIN | | | 0.110 | 0.041 |
| | | | (1.57) | (0.68) |
| EDUC | | | 0.052 | 0.012 |
| | | | (0.67) | (0.17) |
| Constant | 25.300*** | -66.660*** | 27.320 | -61.730*** |
| | (5.13) | (-7.63) | (1.48) | (-3.48) |
| | | | | |
| Industry FE | yes | yes | yes | yes |
| Year FE | yes | yes | yes | yes |
| N | 6,550 | 6,550 | 6,550 | 6,550 |
| adj. $R^2$ | 0.111 | 0.258 | 0.109 | 0.259 |

Table 9 presents results from our multivariate analysis to assess the impact of religious adherence on selective disclosure. *ABS_SD_GW* is a firm's absolute selective disclosure, computed as the absolute difference absolute and weighted disclosure ratio. *RELIGIOSITY* is the number of religious adherents for all denominations in the county where a firm is headquartered divided by the county's total population. *CAPITAL_INTENSITY*, is the ratio between net property, plants and equipment and total assets. *SIZE* is the natural logarithm of firms' net sales. *ROA* is the return on assets. *FOREIGN_SALES* is the percentage of sales to foreign countries. *ENV_DAMAGES* represents total direct environmental costs scaled by total assets. *GOVERNANCE* is Refinitiv's firm governance score. *AGE* is the median age of the county's population in which a firm is headquartered. *EDUC* is calculated as the fraction of population over 25 years that have completed a bachelor's degree or higher. *MALEMIN* is the fraction of male minority population in a county. *POP* is the logarithm of a county's total population. T-statistics are in parentheses. *,**, and *** represent statistical significance at the 10%, 5%, and 1% level, respectively.

Table 10. Transmission channels analysis

| | (1) SD_CHOICE=0 Brownwashing | (2) SD_CHOICE=0 Brownwashing | (3) SD_CHOICE=0 Brownwashing | (4) SD_CHOICE=0 Brownwashing | (5) SD_CHOICE=2 Greenwashing | (6) SD_CHOICE=2 Greenwashing | (7) SD_CHOICE=2 Greenwashing | (8) SD_CHOICE=2 Greenwashing |
|---|---|---|---|---|---|---|---|---|
| RELIGIOSITY | -0.010 | -0.009 | -0.010 | -0.009 | -0.029*** | -0.029*** | -0.027*** | -0.026*** |
| | (-1.29) | (-1.08) | (-1.35) | (-1.06) | (-3.46) | (-3.14) | (-3.12) | (-2.74) |
| FOREIGN_SALES | | 0.033*** | | 0.033*** | | 0.017*** | | 0.016*** |
| | | (8.10) | | (8.05) | | (3.75) | | (3.51) |
| SIZE | | 0.918*** | | 0.914*** | | 1.062*** | | 1.040*** |
| | | (10.59) | | (10.51) | | (10.61) | | (10.37) |
| ROA | | -0.012 | | -0.012 | | 0.009 | | 0.008 |
| | | (-1.01) | | (-1.03) | | (0.60) | | (0.52) |
| CAPITAL_INTENSITY | | 0.024*** | | 0.024*** | | -0.001 | | 0.000 |
| | | (4.57) | | (4.58) | | (-0.20) | | (0.04) |
| ENV_DAMAGES | | 0.162* | | 0.159* | | -0.121 | | -0.121 |
| | | (1.94) | | (1.96) | | (-0.49) | | (-0.50) |
| GOVERNANCE | | 0.024*** | | 0.025*** | | 0.023*** | | 0.023*** |
| | | (7.35) | | (7.52) | | (4.82) | | (4.77) |
| AGE | | | 0.020 | 0.031 | | | -0.013 | -0.007 |
| | | | (0.59) | (0.79) | | | (-0.31) | (-0.16) |
| POP | | | 0.001 | -0.060 | | | 0.118 | 0.052 |
| | | | (0.01) | (-0.55) | | | (0.88) | (0.36) |
| MALEMIN | | | 0.016** | 0.012 | | | 0.011 | 0.002 |
| | | | (2.01) | (1.37) | | | (1.11) | (0.22) |
| EDUC | | | -0.004 | -0.007 | | | 0.019** | 0.014 |
| | | | (-0.47) | (-0.81) | | | (1.96) | (1.38) |
| Constant | 0.079 | -18.760*** | -0.977 | -19.320*** | -2.423*** | -21.060*** | -3.938 | -21.350*** |
| | (0.16) | (-12.01) | (-0.49) | (-6.60) | (-2.62) | (-11.24) | (-1.45) | (-6.74) |
| Industry FE | yes | yes | yes | yes | yes | yes | yes | yes |
| Year FE | yes | yes | yes | yes | yes | yes | yes | yes |
| N | 6,550 | 6,550 | 6,550 | 6,550 | 6,550 | 6,550 | 6,550 | 6,550 |
| pseudo $R^2$ | 0.096 | 0.285 | 0.102 | 0.288 | 0.096 | 0.285 | 0.102 | 0.288 |

Table 10 presents results from multinomial logistic regressions to determine which type of selective disclosure is influenced by religious adherence. SD_CHOICE is a trinary variable which equals 0 if the firm brownwashes, 1 if the firm is not engaged in selective disclosure, and 2 if the firm is engaged in greenwashing. RELIGIOSITY is the number of religious adherents for all denominations in the county where a firm is headquartered divided by the county's total population. CAPITAL_INTENSITY, is the ratio between net property, plants and equipment and total assets. SIZE is the natural logarithm of firms' net sales. ROA is the return on assets. FOREIGN_SALES is the percentage of sales to foreign countries. ENV_DAMAGES represents total direct environmental costs scaled by total assets. GOVERNANCE is Refinitiv's firm governance score. AGE is the median age of the county's population in which a firm is headquartered. EDUC is calculated as the fraction of population over 25 years that have completed a bachelor's degree or higher. MALEMIN is the fraction of male minority population in a county. POP is the logarithm of a county's total population. T-statistics are in parentheses. *,**, and *** represent statistical significance at the 10%, 5%, and 1% level, respectively.

Table 11. Transmission channels analysis

| | (1) GREENWASHING_BIN | (2) GREENWASHING_BIN | (3) GREENWASHING_MAG | (4) GREENWASHING_MAG |
|---|---|---|---|---|
| CATHRT | -0.004 | | -0.026 | |
| | (-0.47) | | (-0.69) | |
| MAINRT | | -0.084*** | | -0.149* |
| | | (-3.22) | | (-1.96) |
| FOREIGN_SALES | -0.001 | -0.003 | -0.022 | -0.026* |
| | (-0.28) | (-0.85) | (-1.45) | (-1.70) |
| SIZE | 0.529*** | 0.545*** | 1.252*** | 1.282*** |
| | (6.77) | (6.92) | (4.47) | (4.60) |
| ROA | 0.012 | 0.008 | 0.045 | 0.040 |
| | (0.88) | (0.61) | (0.96) | (0.87) |
| CAPITAL_INTENSITY | -0.012** | -0.012* | -0.027 | -0.024 |
| | (-2.05) | (-1.93) | (-1.57) | (-1.43) |
| ENV_DAMAGES | -0.288 | -0.264 | -0.217* | -0.247* |
| | (-1.15) | (-1.14) | (-1.66) | (-1.91) |
| GOVERNANCE | 0.009** | 0.009** | 0.024** | 0.023** |
| | (2.24) | (2.22) | (2.09) | (2.00) |
| AGE | 0.000 | -0.022 | 0.006 | -0.046 |
| | (0.01) | (-0.56) | (0.06) | (-0.49) |
| POP | 0.109 | -0.088 | 0.565* | 0.088 |
| | (0.79) | (-0.63) | (1.74) | (0.29) |
| MALEMIN | -0.005 | -0.001 | -0.035 | -0.024 |
| | (-0.49) | (-0.07) | (-1.30) | (-0.89) |
| EDUC | 0.024*** | 0.028*** | 0.153*** | 0.159*** |
| | (2.61) | (2.97) | (3.59) | (3.70) |
| Constant | -13.770*** | -9.894*** | -26.490*** | -17.820*** |
| | (-4.99) | (-3.49) | (-3.98) | (-2.68) |
| Industry FE | yes | yes | yes | yes |
| Year FE | yes | yes | yes | yes |
| N | 6,550 | 6,550 | 6,550 | 6,550 |
| adj. $R^2$ | | | 0.083 | 0.085 |
| pseudo $R^2$ | 0.171 | 0.179 | | |

Table 11 presents results from our multivariate analysis to assess the impact of catholic and protestant religious adherence on greenwashing likelihood and greenwashing magnitude. Models (1) and (2) are logistic regressions. Models (3) and (4) are ordinary least square regressions. GREENWASHING_BIN is a binary variable that equals one if a firm greenwashes and zero otherwise. GREENWASHING_MAG is the difference between the absolute disclosure ratio and the weighted disclosure ratio if a firm's absolute disclosure ratio is strictly above its weighted disclosure ratio, and zero otherwise. CATHRT (columns 1 and 3) is the ratio between Catholic adherents in the county where a firm's headquarter is located divided by the county's total population. MAINRT (column 2 and 4) is the ratio between Protestant adherents in the county where a firm's headquarter is located divided by the county's total population. CAPITAL_INTENSITY, is the ratio between net property, plants and equipment and total assets. SIZE is the natural logarithm of firms' net sales. ROA is the return on assets. FOREIGN_SALES is the percentage of sales to foreign countries. ENV_DAMAGES represents total direct environmental costs scaled by total assets. GOVERNANCE is Refinitiv's firm governance score. AGE is the median age of the county's population in which a firm is headquartered. EDUC is calculated as the fraction of population over 25 years that have completed a bachelor's degree or higher. MALEMIN is the fraction of male minority population in a county. POP is the logarithm of a county's total population. T-statistics are in parentheses. *,**, and *** represent statistical significance at the 10%, 5%, and 1% level, respectively.